\documentclass[12pt,preprint]{aastex}

\begin{document}
\title{A Ground-Based Search for Thermal Emission from the Exoplanet TrES-1}
\author{Heather A. Knutson, David Charbonneau\altaffilmark{1}}
\affil{Harvard-Smithsonian Center for Astrophysics, 60 Garden Street, Cambridge, MA 02138}
\altaffiltext{1}{Alfred P. Sloan Research Fellow}
\email{hknutson@cfa.harvard.edu, dcharbonneau@cfa.harvard.edu}

\author{Drake Deming}
\affil{Planetary Systems Laboratory, Code 693, Goddard Space Flight Center, Greenbelt, MD 20771}
\email{ddeming@pop600.gsfc.nasa.gov}

\and
\author{L. Jeremy Richardson}
\affil{Exoplanet and Stellar Astrophysics Laboratory, Code 667, Goddard Space Flight Center, Greenbelt, MD 20771}
\email{richardsonlj@milkyway.gsfc.nasa.gov}

\begin{abstract}

Eclipsing planetary systems give us an important window on extrasolar planet atmospheres.  By measuring the depth of the secondary eclipse, when the planet moves behind the star, we can estimate the strength of the thermal emission from the day side of the planet.  Attaining a ground-based detection of one of these eclipses has proven to be a significant challenge, as time-dependent variations in instrument throughput and atmospheric seeing and absorption overwhelm the small signal of the eclipse at infrared wavelengths.  We gathered a series of simultaneous $L$ grism spectra of the transiting planet system TrES-1 and a nearby comparison star of comparable brightness, allowing us to correct for these effects in principle.  Combining the data from two eclipses, we demonstrate a detection sensitivity of $0.15\%$ in the eclipse depth relative to the stellar flux.  This approaches the sensitivity required to detect the planetary emission, which theoretical models predict should lie between $0.05-0.1\%$ of the stellar flux in our $2.9-4.3$~\micron~bandpass.  We explore the factors that ultimately limit the precision of this technique, and discuss potential avenues for future improvements.
                                                        
\end{abstract}


\section{Introduction}\label{intro}

Recent observations using the \emph{Spitzer Space Telescope} have detected the thermal emission from several transiting planets, including TrES-1\citep{dem05,dem06,char05,grill07,rich07,swain07}.  By measuring the decrease in flux when the planet moves behind the star, \citet{char05} estimated the brightness temperature of the planet at  4.5~\micron~and 8.0~\micron.  In the shorter wavelength bandpass, they estimated an eclipse depth of $0.00066\pm0.00013$ in relative flux for TrES-1, corresponding to a brightness temperature of $1010\pm60$~K for the planet.  This brightness temperature is significantly lower than predicted by models, which indicate that hot Jupiters should have a strong peak in their emission at 4 \micron, created by a gap between absorption bands from CO and H$_2$O.  There are a number of possible explanations for the low flux at 4.5~\micron, including a higher-than-expected abundance of CO and additional opacity sources in the atmosphere, but the bandpass-integrated nature of \emph{Spitzer} photometry make it difficult to determine the correct explanation.   

In this paper we describe a series of $L$-band spectroscopic observations of two secondary eclipses of TrES-1 \citep{alonso04,sozz04,laugh05,winn07} using the NIRI instrument on Gemini North.  Because the ultimate goal is to constrain theoretical model spectra, spectroscopic observations are inherently more useful; although we must still bin our data over a range of wavelengths in order to detect the secondary eclipse, we could, in principle, vary our bandpasses to measure the amplitude of the 4~\micron~peak and other spectral features directly.  For our measurement we observe TrES-1 and a close comparison star simultaneously in the same slit and use this second set of spectra to remove the effects of variable atmospheric absorption and slit losses due to telescope pointing jitter and seeing. 

Our data represent the first ground-based attempt to detect the secondary eclipse of TrES-1; previous ground-based studies \citep{rich03a,rich03b,snell05,dem07} focused on HD 209458b, which lacks a nearby comparison star of similar infrared brightness.  As a result, \citet{rich03a,rich03b},\citet{dem07} and \citet{snell05} were forced to nod between the target and the calibration star, which limited their ability to adequately remove these time-dependent variations.  \citet{dem07} note that they achieved the most accurate correction using the nearer of their two comparison stars, at a separation of 1.7\degr, but the lack of coherence between the observed fluxes for the two stars indicated that absorption from telluric water vapor was still incoherent on this angular scale.  \citet{snell07} obtained relative photometry spanning secondary eclipse for the transiting planet system OGLE-TR-113, which is located in a crowded field.  The 1.3\arcmin~square field of view contained several nearby stars of comparable brightness for calibration.  They report a tentative detection of the secondary eclipse, indicating that the use of nearby comparison stars can significantly reduce errors.  Similarly, we selected TrES-1 for our experiment because it has a nearby comparison star; by placing both stars in the slit simultaneously, we planned to simply ratio the two observed spectra to correct for time-dependent instrumental and telluric variations.

\section{Observations and Analysis}\label{methods}

We used the Near InfraRed Imager (NIRI) on Gemini North \citep{hod03} to observe TrES-1 and a nearby star (2MASS 19041058+3638409, 44\arcsec~distant) in the same slit during secondary eclipses on UT 2006 May 11 and July 26.  We obtained a series of 758 $L$ spectra of each star spanning the wavelength range from $2.9-4.3$~\micron.  We chose the widest slit available for $L$ grism spectroscopy, with dimensions of $0.75\arcsec \times 110\arcsec$, to maximize the total flux and minimize slit losses due to guiding errors.  We used 3~s exposures, with 15~coadds each, for our images.  

\subsection{Image Reduction}\label{image_red}

In order to remove the sky background we nod the telescope in an ABBA pattern by by $\pm5\arcsec$ for the May eclipse and $\pm2.5\arcsec$ for the July eclipse, then difference the resulting pairs of images.  Immediately after differencing the two images we fit the data with a function designed to remove periodic detector noise (see the end of \S\ref{decorrel} for a full description of this step).  Because the flats were taken with the shutter closed they are dominated by the thermal blackbody spectrum of the warm shutter.  We remove this effect by taking the central 200 rows in the flat-field image and binning them to make a spectrum, then divide all rows in our flat-field image by this composite spectrum.  After flat-fielding the data, we calculate the errors associated with each pixel in our images using the equation derived by \citet{vacca04}.  The variance in electron counts associated with an individual pixel in the images before flat-fielding is given by:
\begin{equation}\label{err_eq}
V_I=gI + 2n_c\sigma^2_{read}
\end{equation}
where $I$ is the flux in $ADU$, $g$ is the gain ($g=12.3$~$e^-$~$ADU^{-1}$ for these images), $n_c$ is the number of co-adds, and $\sigma_{read}$ is the rms read noise ($\sigma_{read}=70$~$e^-$~$pix^{-1}$).  Once the raw images are differenced and flat-fielded, the per pixel variance is given by:
\begin{equation}\label{comb_err}
V_D=\frac{V_A+V_B}{flat^2}
\end{equation}
where $V_A$ is the variance for the image at position A, $V_B$ is the variance for position B and $flat$ is the flat-field image.

\subsection{Creating the Timeseries}\label{create_timeseries}

\begin{figure}
\epsscale{0.7}
\plotone{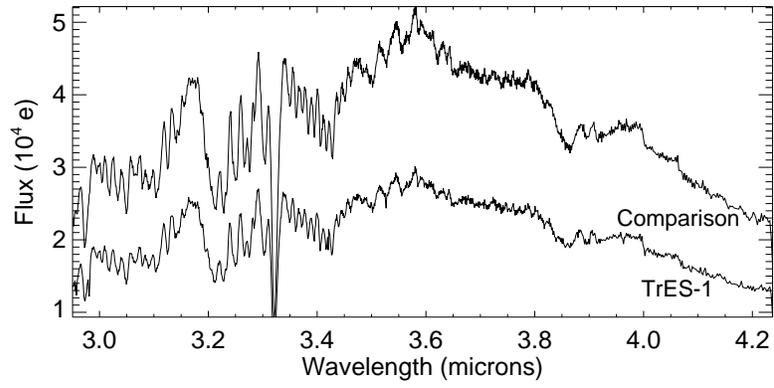}
\caption{These are the averaged spectra for TrES-1 (bottom) and the companion star (top) from UT 2006 July 26.  The dominant spectral features are from telluric absorption.\label{spectra}}
\end{figure}

\begin{figure}
\epsscale{0.7}
\plotone{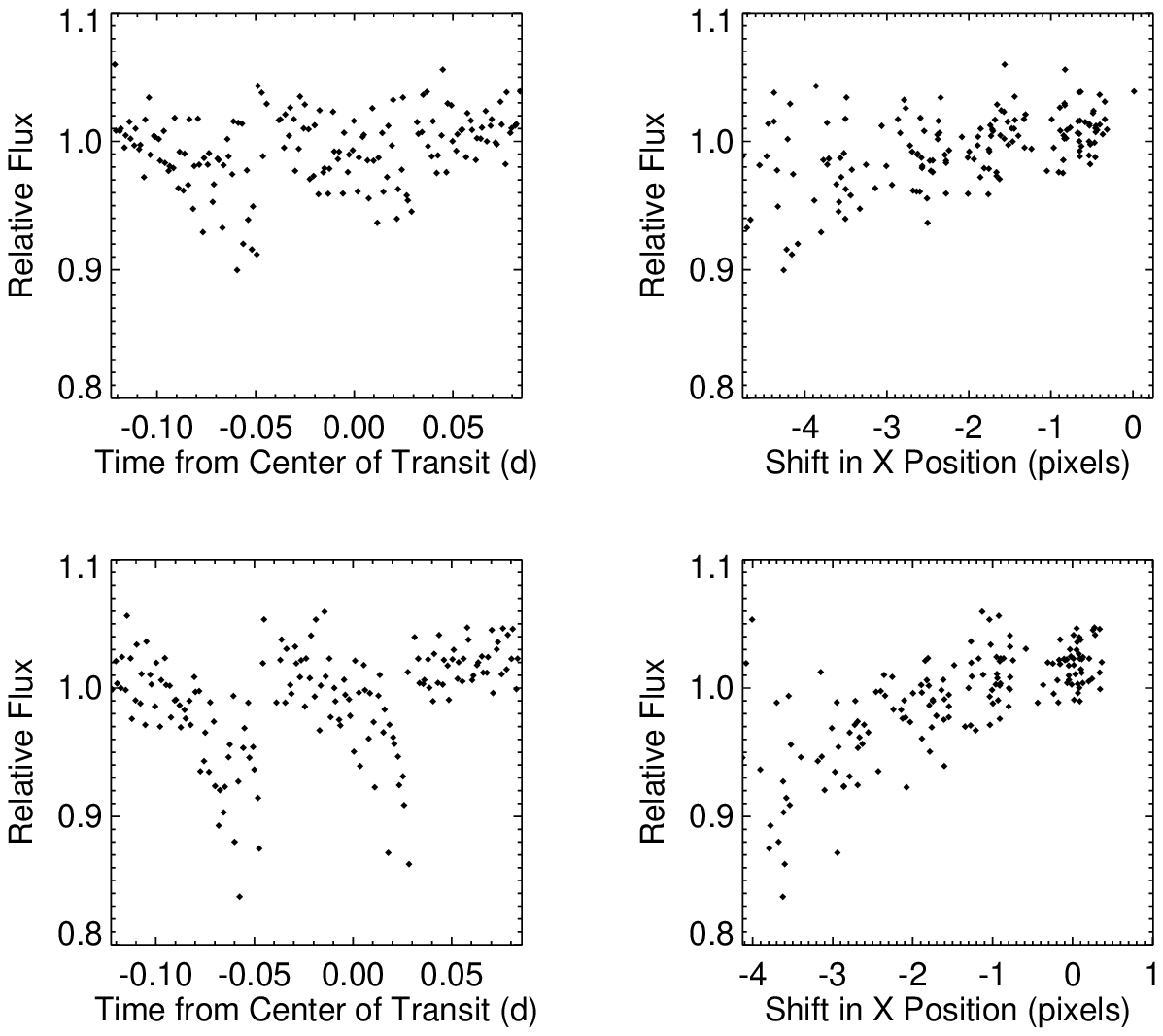}
\caption{The panels on the left show the relative flux values as a function of time for the UT 2006 May 11 eclipse, for nod positions A (top) and B (bottom), and the panels on the right show the correlation between these values and shift in the $x$ (dispersion) direction.\label{May_correl}}
\end{figure}

\begin{figure}
\epsscale{0.7}
\plotone{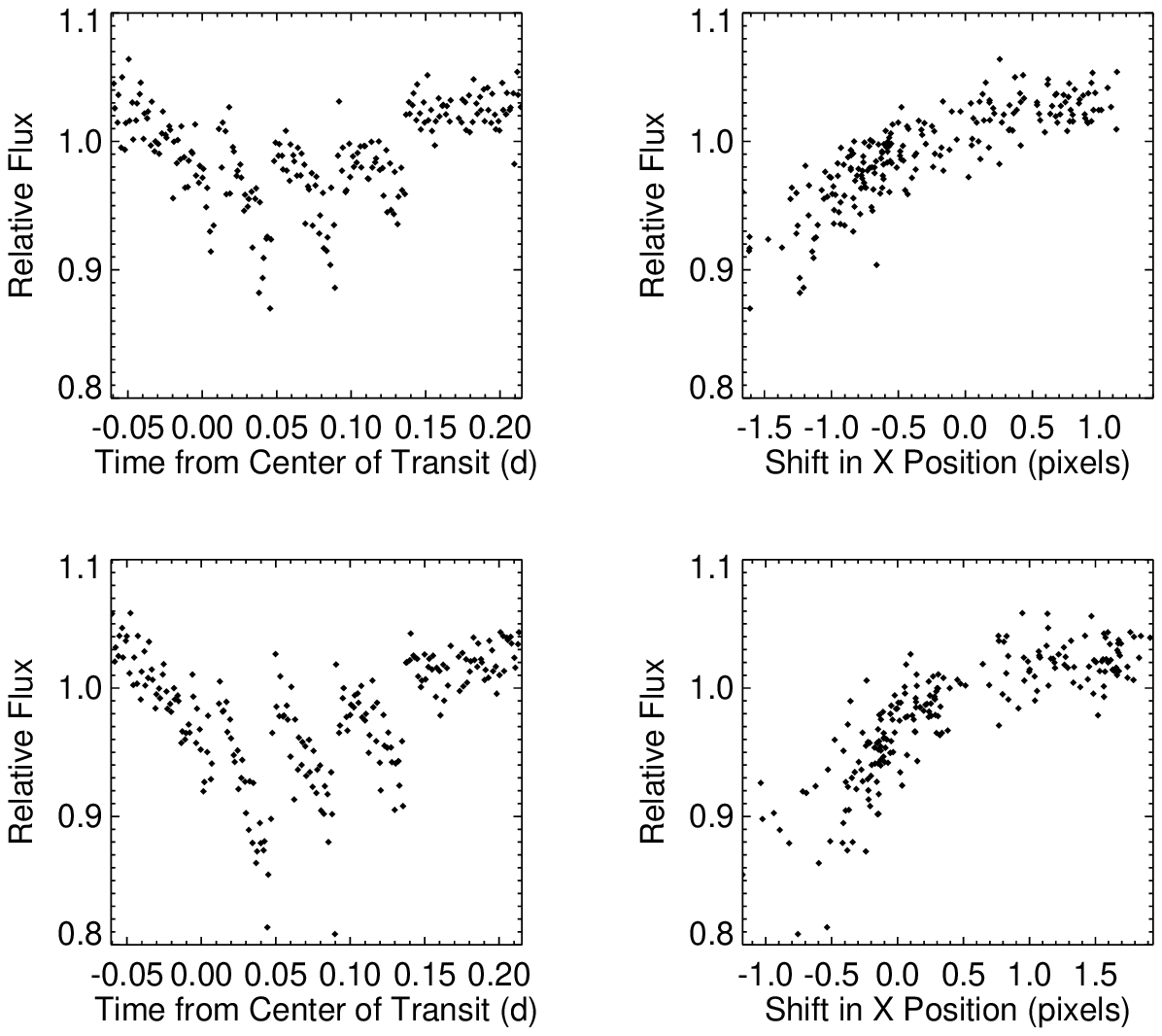}
\caption{The panels on the left show the relative flux values as a function of time for the UT 2006 July 26 eclipse, for nod positions A (top) and B (bottom), and the panels on the right show the correlation between these values and the shift in the $x$ (dispersion) direction.\label{July_correl}}
\end{figure}

We use optimal spectral extraction \citep{horne86} to extract the one-dimensional spectra from our normalized images, following the method outlined by \citet{cush04} and treating each of the four spectra in the differenced image separately.  This gives us a total of 758 spectra for TrES-1 and 758 spectra for the companion star (see Figure \ref{spectra}).  From this point on we treat data from each of the two nights and each of the two nod positions as independent data sets; as shown in Figures \ref{May_correl} and \ref{July_correl}, each of these data sets has different noise properties.  We attempt to correct for instrumental variations as follows:  First, we calculate a median spectrum for the comparison star and divide all of the spectra from the comparison star by this median spectrum.  This effectively flattens the spectra, removing the constant component in their shape and leaving the relative changes in flux at each wavelength over time.  We then divide each spectrum of TrES-1 by the corresponding flattened spectrum from the companion star, thus removing any temporal variations that are common to both stars.

Next we bin each of the normalized spectra of TrES-1 over the entire wavelength range to create a timeseries.  We chose to exclude wavelengths that have negative flux values in any of the spectra from a given night, as these values indicate either (a) a column of bad pixels or (b) low fluxes.  Because our images are dominated by the sky background, regions with low flux will generally contribute more noise than signal to the final timeseries, and excluding them reduces the RMS variation in the binned timeseries.  We exclude 125 of 1024 wavelength pixels for the UT 2006 May 11 eclipse, and 63 wavelength pixels for the UT 2006 July 26 eclipse.  Most of the excluded pixels were from either the 100 reddest pixels or the 20 bluest pixels, where atmospheric absorption is strongest.  We then normalize the data by dividing the timeseries by the median out-of-transit value.  The resulting timeseries are shown in Fig. \ref{May_correl} and \ref{July_correl}.

\subsection{Removing Systematic Trends}\label{decorrel}

Variations at the level of $10-20\%$ are apparent in the timeseries.  These variations are well-correlated with the offsets of the spectra in the $x$ (dispersion) direction, which serves as a proxy for the position of the star along the short axis of the slit.  We estimate these offsets by cross-correlating the strong telluric feature around 3.3~\micron~ in individual spectra with a composite spectrum made by combining ten individual spectra gathered at low airmass (this ensures our template spectrum has a higher signal to noise than the individual spectra).  We attempt to remove these trends by fitting the out-of-eclipse data at each nod position separately with a quadratic function of the $x$ shift.  Additional degrees of freedom did not produce significant improvements.  This decorrelation reduces the level of variation to $1-2\%$ (see Figure \ref{boxcar}).  We searched for additional correlations with variables including airmass, detector temperature, $y$ offset (cross-dispersion direction), various proxies for changes in focus including the width of the dispersion profile of the 2D spectra and the width of the cross-correlation function of the extracted 1D spectra, and the total flux from the calibration star.  None of these variables correlate with the residuals, although the small size of these trends relative to the scatter in the data makes it difficult to rule out subtle correlations with a high degree of confidence.  

The total binned fluxes from the two stars varied by a factor of three during our observations as the stellar centroids moved in and out of the slit center.  Although most of this variation is removed when we take the ratio of the TrES-1 spectrum to the companion star's spectrum, it is possible that a nonlinear detector response might account for some of the residual trends.  Our raw images, which are dominated by sky emission, have a typical flux ranging from $1\times10^4$~$e^-$ at the shortest wavelengths to $8\times10^4$~$e^-$ at the longest wavelengths.  Although this is well below the saturation limit of $3\times10^6$~$e^-$, the instrument is still nonlinear at a level of approximately 1\% for these illumination levels (Andrew Stephens, personal correspondence, Sept. 2006).  To quantify the effect that this nonlinearity might have on our final binned timeseries, we apply the following nonlinearity correction to our raw images.  We assumed the nonlinearity scaled with flux, according to:
\begin{equation}
f^1=f\left(1+K\left(\frac{f}{6\times10^4~e^-}\right)\right)
\end{equation}
where $f$ is the original flux in $e^-$ and $f^1$ is the corrected flux, and $K$ is a coefficient describing the nonlinearity at $6\times10^4$$e^-$.  We assumed values of $K=0.01,0.03,0.05$ and $0.15$.  When we created our final timeseries using these corrected images, individual values in the normalized timeseries changed by only $0.03\%-0.07\%$ on average and the overall variance in the data was effectively unchanged.

We note that the RMS variation of the decorrelated timeseries exceeds the expectations from Poisson and read noise by a factor of 4. It is possible that some of this noise might be related to the properties of the detector;  infrared detectors often have signficant power close to the Nyquist frequency.  Indeed, when we examined our images in Fourier space we noticed there was significant row-to-row power at a period of two pixels.  Although most of this noise is removed when the images are pair-subtracted, a small but significant peak remains in the power spectrum at a frequency of 0.25 $pix^{-1}$.

To remove these residuals, we treat each quadrant in the pair-subtracted images separately.  We fit each column with a sine function with a fixed period of four pixels and solve for the phase and the amplitude, which was permitted to vary quadratically as a function of position.  We then subtract the best-fit function from the column, and repeat for all the columns in the quadrant.  This method removed 98\% of the power at four-pixel frequencies.  When we created our timeseries using images with four-pixel frequencies removed, the flux values of individual points changed by 0.15\% on average and the overall variance in the data was unchanged.

\section{Discussion}\label{eclipse_lim}

\begin{figure}
\epsscale{0.7}
\plotone{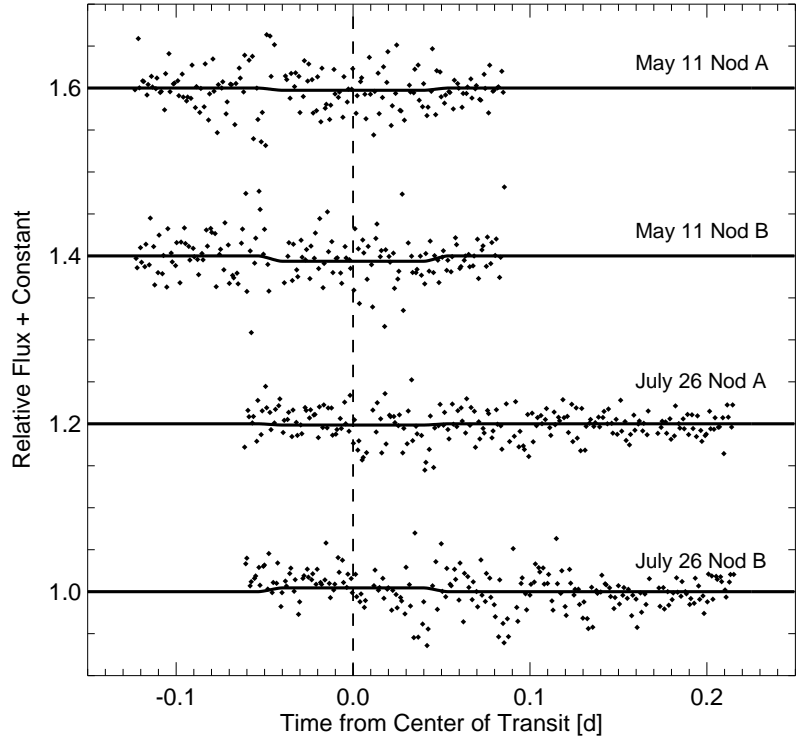}
\caption{The relative change in flux for both eclipses and nod positions, along with best-fit transit curves for each timeseries.  We treat each of the two nod positions separately in our analysis, as they have different dependencies on the shift in the $x$ (dispersion) direction and airmass.\label{boxcar}}
\end{figure}

For each of the four timeseries, we estimate the depth of the eclipse (Table \ref{eclipse_depths}).  We fix the system parameters to the best-fit values published by \citet{winn07}, and allowed the depth of the eclipse to vary over both positive and negative values.  We calculated our errors using a bootstrap Monte Carlo analysis.  We note that the only time series to give a positive eclipse depth (July 26 Nod B) also contains the largest correlated residuals (Fig \ref{boxcar}).  When we take the weighted average of these estimates, we find an eclipse depth of $-0.0010\pm0.0015$ in relative flux, which is consistent with no variation.  We also note that the eclipse depth may vary over time due to changing emission patterns on the planet \citep{rau07}, although this variation is predicted to be much smaller than the sensitivity of our measurements.  Published models of the emitted dayside spectrum for hot Jupiters \citep{fort05,seag05,bar05,bur06} predict that the depth of the eclipse at the wavelengths of our observations will depend on the strength of various molecular absorption bands, including CO and H$_2$O, as well as the efficiency with which energy is circulated between the permanent day and night sides of the planet.  \citet{fort05} predict eclipse depths for TrES-1 of $-0.0010$ to $-0.0005$ for this bandpass, with the smaller eclipse depths corresponding to a model in which energy is efficiently recirculated from the dayside to the nightside in the planetary atmosphere, and the larger eclipse depths corresponding to inefficient circulation and a large day-night temperature differential.

We also considered a restricted bandpass from $3.8-4.1$~\micron, centered on the 4~\micron~peak in the planet's emission predicted by theoretical models. \citet{fort05} predict that TrES-1 should have an eclipse depth between $-0.0015$ and $-0.0009$ for this wavelength range.  We estimate a value of $+0.0061\pm0.0026$ (i.e. an increase in flux during the eclipse) for this wavelength range.  The trends that are visible in Figure \ref{boxcar} are even larger for these smaller wavelength ranges, thus we conclude that the measurement is not significant.  We also binned the spectrum into a red and blue timeseries and took the ratio of these two timeseries to search for color-dependent variations in the depth of the eclipse, but this increased the overall noise level and failed to remove any of the systematic trends, which are different in these two regions of the spectrum.

We have not been able to determine the precise source of the noise that ultimately limits our measurements, but there are several conclusions that we draw from the analysis.  First, there do not appear to be any significant correlations between relative flux variations and airmass, humidity, temperature, and other properties of the local observing conditions.  This is an improvement over previous studies \citep{rich03a,rich03b,snell05,snell07,dem07}, which concluded that the correction for atmospheric absorption was the limiting factor when the comparison star was located at distances of 2\arcmin~or more.

A central limitation to our experimental design is the slit, which when coupled to pointing jitter and variations in the focus and seeing introduces large changes in the apparent flux.  In particular, we see a strong correlation between the centroid of the stars on the slit and the measured flux levels, even after dividing the measured flux for TrES-1 by the companion's flux.  Although we are able to remove most of these variations by decorrelating with the $x$ position of the spectra, there are still some trends remaining after our decorrelation.   We note that most of the issues listed above would be mitigated with the use of a wider slit; although we selected the widest slit available on NIRI, this slit width was still comparable to the point spread functions of the two stars and required us to make multiple adjustments of the telescope pointing over the course of the night to counteract the drift of stars out of the slit.  Because we must bin the data over a relatively wide range in wavelength to reduce the photon noise to the level required to seek the secondary eclipse, the loss in wavelength resolution that would result from a wider slit would not be significant.  Although spectroscopic observations of the secondary eclipse are scientifically more valuable, we note that photometric observations would also avoid many of the problems encountered here.  Previous ground-based attempts to detect the secondary eclipse of HD 209458 were limited by the lack of nearby bright stars in the field of view; it is entirely possible that these observations could be successful if they were repeated with targets such as TrES-1, which have such nearby companions.

\section{Conclusions}\label{conclusions}

We estimate a depth of $-0.0010\pm0.0015$ for the eclipse from $2.9-4.3$~\micron, which is consistent with zero.  Despite employing a method designed to remove time-dependent variations in atmospheric absorption and seeing, we were unable to reduce the noise to the level required to address theoretical models of the planetary emission at these wavelengths.  The noise in our data is significantly higher than that predicted from Poisson and detector read noise.  This may be related to the interplay between seeing and slit losses, although we were unable to verify this directly.  Based on our experience with these data, it is possible to accurately correct for time-dependent variations in atmospheric absorption for stars with nearby (less than 1\arcmin) bright companions, and remaining issues with slit losses may be addressed with either a wider slit or purely photometric (as opposed to spectroscopic) observations.  Although our proposed modifications would reduce our ability to resolve features in the emission spectra of these planets, such compromises may be necessary in order to achieve a successful detection of a secondary eclipse from the ground.  This will become increasingly important in the next several years, as \emph{Spitzer} is predicted to run out of cryogen in early 2009.  Additionally, ground-based transit surveys \citep{alonso04,mcc06,odon06,coll07,bak07} are expected to double the number of known transiting planet systems by the end of 2007.  These surveys tend to target bright stars in relatively crowded fields, and it is likely that many of these newly discovered systems will have bright nearby comparison stars that make them suitable for the kinds of observations we have described above.

\acknowledgements

This work is based on observations obtained as part of program GN-2006A-Q-3 at the Gemini Observatory, which is operated by the Association of Universities for Research in Astronomy, Inc., under a cooperative agreement with the NSF on behalf of the Gemini partnership: the National Science Foundation (United States), the Particle Physics and Astronomy Research Council (United Kingdom), the National Research Council (Canada), CONICYT (Chile), the Australian Research Council (Australia), CNPq (Brazil) and CONICET (Argentina).  We are grateful to Chad Trujillo and the entire Gemini team for their assistance throughout this process. HAK was supported by a National Science Foundation Graduate Research Fellowship. LJR was supported by a NASA Postdoctoral Fellowship at NASA Goddard.

\clearpage

\begin{deluxetable}{lrrrrcrrrrr}
\tabletypesize{\scriptsize}
\tablecaption{Estimates of the eclipse depth ($2.9-4.3$~\micron)\label{eclipse_depths}}
\tablewidth{0pt}
\tablehead{
\colhead{Date} & \colhead{Depth at Position A}  & \colhead{Depth at Position B}}
\startdata
UT 2006 May 11 & $-0.0026\pm0.0033$ & $-0.0063\pm0.0040$ \\
UT 2006 July 26 & $-0.0015\pm0.0024$ & $+0.0045\pm0.0031$ \\
\enddata
\end{deluxetable}

\clearpage

\clearpage

\clearpage

\clearpage

\end{document}